\documentclass[aps,prl,reprint,amsmath,amsfonts,amssymb]{revtex4-1}
\usepackage{adjustbox,graphicx,bm,color,microtype,hyperref,physics}
\usepackage{tikz}
\usetikzlibrary{decorations.pathreplacing,decorations.pathmorphing,arrows,backgrounds,fit,shapes}

\newcommand{\eps}{\varepsilon}
\newcommand{\rs}{r_s}
\newcommand{\rsSB}{r_s^\text{SB}}
\newcommand{\mc}{\multicolumn}
\newcommand{\eFF}{e_\text{FF}}
\newcommand{\tFF}{t_\text{FF}}
\newcommand{\vFF}{v_\text{FF}}
\newcommand{\eWC}{e_\text{WC}}
\newcommand{\vWC}{v_\text{WC}}
\newcommand{\tWC}{t_\text{WC}}

\newcommand{\pFF}{\rho_{\text{FF}}}
\newcommand{\pWC}{\rho_{\text{WC}}}
\newcommand{\pGSWC}{\rho_{\text{GSWC}}}
\newcommand{\pCDW}{\Delta \rho_{\text{CDW}}}
\newcommand{\kF}{k_\text{F}}
\newcommand{\eF}{\eps_\text{F}}
\newcommand{\mL}{\mathcal{L}}
\newcommand{\mR}{\mathcal{R}}

\begin{document}

\title{Excited-State Wigner Crystals in One Dimension}

\author{Fergus J. M. Rogers}
\affiliation{Research School of Chemistry, Australian National University, Canberra ACT 2601, Australia}
\author{Pierre-Fran{\c c}ois Loos}
\email[Corresponding author: ]{pf.loos@anu.edu.au}
\affiliation{Research School of Chemistry, Australian National University, Canberra ACT 2601, Australia}
\begin{abstract}
Wigner crystals (WC) are electronic phases peculiar to low-density systems, particularly in the uniform electron gas.
Since its introduction in the early twentieth century, this model has remained essential to many aspects of electronic structure theory and condensed-matter physics.  
Although the (lowest-energy) ground-state WC (GSWC) has been thoroughly studied, the properties of excited-state WCs (ESWCs) are basically unknown.
To bridge this gap, we present a well-defined procedure to obtain an entire family of ESWCs in a one-dimensional electron gas using a symmetry-broken mean-field approach.
While the GSWC is a commensurate crystal (i.e.~the number of density maxima equals the number of electrons), these ESWCs are incommensurate crystals exhibiting more or less maxima.
Interestingly, they are lower in energy than the (uniform) Fermi fluid state.
For some of these ESWCs we have found asymmetrical band gaps, which would lead to anisotropic conductivity.
These properties are associated to unusual characteristics in their electronic structure.
\end{abstract}


\maketitle

\textit{Wigner crystals.---}
In 1934, Wigner predicted that, at low density, electrons within a positively-charged uniform background (or jellium) would ``crystallize" onto lattice sites, thus forming electronic analogues of the well-known atomic crystals \cite{Wigner34}.
As opposed to the delocalized, uniform Fermi fluid (FF) state that minimizes the kinetic energy (whose contribution is predominant for high densities), these exotic phases, known as Wigner crystals (WCs), curtail the interelectronic Coulomb interaction, which dominates at low densities \cite{VignaleBook}.
The morphology of a WC is made apparent by the occurrence of periodic maxima or ``peaks'' in the electron density.
The most common crystal symmetries are the bcc, fcc and hcp lattices in three dimensions (3D), the triangular and square lattices in two dimensions (2D), and the evenly-spaced lattice in one dimension (1D) \cite{UEG2016}.

WCs have received renewed interest in recent literature, particularly in the uniform electron gas (UEG) or jellium, where they play a central role in the phase diagram \cite{UEG2016}.
Similarly, low-dimensional WCs have also come under scrutiny theoretically and experimentally, as a paradigm for quasi-1D materials \cite{Schulz93, Fogler05a}, such as carbon nanotubes \cite{Bockrath99, Ishii03, Deshpande08} or nanowires \cite{Meyer09, Deshpande10}.

As a consequence of the intensive and ongoing investigation into the ground-state properties of WCs, little is known about their excited states, aside from the plain distinction that some solutions are lower in energy than others.
Recently, we have shown that the ground-state WC (GSWC) is always commensurate in the 1D electron gas (1DEG) \cite{SBLDA1D16}. 
However, using self-consistent field (SCF) Hartree-Fock (HF) calculations, we have since observed the appearance of excited-state WCs (ESWCs), whose periodicities are inherit to the charge-density waves (CDWs) from which they evolved. 
Here, we attribute the GSWC to an $n$-peak WC, while an ESWC is a WC whose peak count $p$ deviates from the number of electrons $n$ by an integer amount. 
These ESWCs can be regarded as two varieties of incommensurate crystals, where the number of peaks $p$ is smaller (supersaturated crystals) or larger (unsaturated crystals) than $n$.
Thus, we propose in this paper to study ESWCs in a 1DEG, and how they relate to the GSWC and the FF. 
The $p > n$ crystals are particularly compelling given that such solutions are known to be the HF ground state of 2D and 3D electron gases at high densities \cite{Zhang08, Bernu08, Bernu11, Baguet13, Baguet14, Delyon15}.
Atomic units are used throughout.

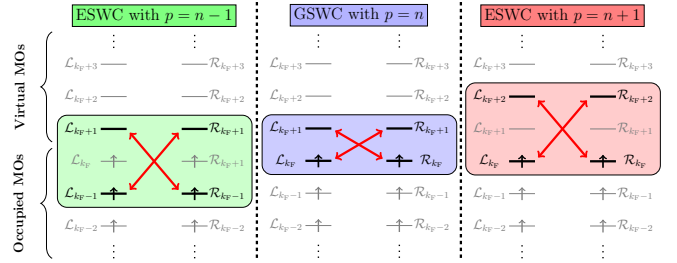
\begin{figure}
\begin{adjustbox}{max width=\linewidth}
\begin{tikzpicture}[scale=1,
    shell3/.style={rectangle,minimum size=6mm,fill=green!50!white,draw=black},
    mixarrow/.style={<->,red,line width=2pt}]
    
    \foreach \x in {-2,0}
        \draw[black!50!white] (-0.9,\x) -- (-0.1,\x)
          (1.6,\x) -- (2.4,\x);%
    \foreach \x in {2,3}
        \draw[black!50!white] (-0.9,\x) -- (-0.1,\x)
          (1.6,\x) -- (2.4,\x);%
    \draw[line width = 2pt] (-0.9,1) -- (-0.1,1)
          (1.6,1) -- (2.4,1)
          (-0.9,-1) -- (-0.1,-1)
          (1.6,-1) -- (2.4,-1);%
          
    \foreach \x in {-2,0}
        \draw[->,very thick,black!50!white] (-0.5,\x-0.2) -- (-0.5,\x+0.2);%
    \foreach \x in {-2,0}
        \draw[->,very thick,black!50!white] (2.0,\x-0.2) -- (2.0,\x+0.2);%
    \draw[->,line width=1.5pt] (-0.5,-1.2) -- (-0.5,-0.8);%
    \draw[->,line width=1.5pt] (2.0,-1.2) -- (2.0,-0.8);%
    
    \draw[loosely dotted, line width=1.2pt] 
        (-0.5,-3.0) -- (-0.5,-2.5)
        (2.0,-3.0) -- (2.0,-2.5)
        (-0.5,3.5) -- (-0.5,4.0)
        (2.0,3.5) -- (2.0,4.0);%
                      
    \node [black!50!white] (lh) at (-1.5,0) {\large $\mL_{\kF}$};%
    \node [black!50!white] (rh) at (3.0,0) {\large $\mR_{\kF+1}$};%
    \node (ll) at (-1.5,1) {\large $\mL_{\kF+1}$};%
    \node (rl) at (3.0,1) {\large $\mR_{\kF+1}$};%
    \node (lh1) at (-1.5,-1) {\large $\mL_{\kF-1}$};%
    \node (rh1) at (3.0,-1) {\large $\mR_{\kF-1}$};%

    \foreach \x in {2,...,2}
        \node [black!50!white] at (-1.5,-\x) {\large $\mL_{\kF-\x}$};%
    \foreach \x in {2,...,2}
        \node [black!50!white] at (3.0,-\x) {\large $\mR_{\kF-\x}$};%
    \foreach \x in {2,...,3}
        \node [black!50!white] at (-1.5,\x) {\large $\mL_{\kF+\x}$};%
    \foreach \x in {2,...,3}
        \node [black!50!white] at (3,\x) {\large $\mR_{\kF+\x}$};%

    \node [shell3]   (n-1)  at  (0.75,4.5)  {\Large ESWC with $p=n-1$};%

\begin{scope}[on background layer]
    \node [fill=green!20!white!,rounded corners=3mm,draw=black,fit=(lh1) (rh1) (ll) (rl)] {};%
\end{scope}%

    \node (ll) at (-0.1,1) {};%
    \node (rl) at (1.6,1) {};%
    \node (lh1) at (-0.1,-1) {};%
    \node (rh1) at (1.6,-1) {};%

    \draw[mixarrow] (ll)--(rh1);%
    \draw[mixarrow] (lh1)--(rl);%

    \draw[thick,decorate,decoration={brace,amplitude=10pt}]
  (-2.4,0.6) -- node[xshift=-1.0cm,rotate=90] {\large \textbf{Virtual MOs}} (-2.4,4.0);%
    \draw[thick,decoration={brace,amplitude=10pt},decorate]
  (-2.4,-3.0) -- node[xshift=-1.0cm,rotate=90] {\large \textbf{Occupied MOs}} (-2.4,0.4);%
	
\end{tikzpicture}
\begin{tikzpicture}[scale=1,
    shell1/.style={rectangle,minimum size=6mm,fill=blue!30!white,draw=black},
    mixarrow/.style={<->,red,line width=2pt}]
    
    \foreach \x in {-2,...,0}
        \draw[black!50!white] (-0.9,\x) -- (-0.1,\x)
          (1.6,\x) -- (2.4,\x);%
    \foreach \x in {2,3}
        \draw[black!50!white] (-0.9,\x) -- (-0.1,\x)
          (1.6,\x) -- (2.4,\x);%
    \draw[line width = 2pt] (-0.9,1) -- (-0.1,1)
          (1.6,1) -- (2.4,1)
          (-0.9,0) -- (-0.1,0)
          (1.6,0) -- (2.4,0);%
          
    \foreach \x in {-2,-1}
        \draw[->,very thick,black!50!white] (-0.5,\x-0.2) -- (-0.5,\x+0.2);%
    \foreach \x in {-2,-1}
        \draw[->,very thick,black!50!white] (2.0,\x-0.2) -- (2.0,\x+0.2);%
    \draw[->,line width=1.5pt] (-0.5,-0.2) -- (-0.5,0.2);%
    \draw[->,line width=1.5pt] (2.0,-0.2) -- (2.0,0.2);%
    
    \draw[loosely dotted, line width=1.2pt] 
        (-0.5,-3.0) -- (-0.5,-2.5)
        (2.0,-3.0) -- (2.0,-2.5)
        (-0.5,3.5) -- (-0.5,4.0)
        (2.0,3.5) -- (2.0,4.0);%
                      
    \node (lh) at (-1.5,0) {\large $\mL_{\kF}$};%
    \node (rh) at (3.0,0) {\large $\mR_{\kF}$};%
    \node (ll) at (-1.5,1) {\large $\mL_{\kF+1}$};%
    \node (rl) at (3.0,1) {\large $\mR_{\kF+1}$};%

    \foreach \x in {1,...,2}
        \node [black!50!white] at (-1.5,-\x) {\large $\mL_{\kF-\x}$};%
    \foreach \x in {1,...,2}
        \node [black!50!white] at (3.0,-\x) {\large $\mR_{\kF-\x}$};%
    \foreach \x in {2,...,3}
        \node [black!50!white] at (-1.5,\x) {\large $\mL_{\kF+\x}$};%
    \foreach \x in {2,...,3}
        \node [black!50!white] at (3,\x) {\large $\mR_{\kF+\x}$};%

    \node [shell1]    (n)     at  (0.75,4.5)      {\Large GSWC with $p=n$};%

\begin{scope}[on background layer]
    \node [fill=blue!20!white!,rounded corners=3mm,draw=black,fit=(lh) (rh) (ll) (rl)] {};%
\end{scope}%

    \node (ll) at (-0.1,1) {};%
    \node (rl) at (1.6,1) {};%
    \node (lh) at (-0.1,0) {};%
    \node (rh) at (1.6,0) {};%

    \draw[mixarrow] (ll)--(rh);%
    \draw[mixarrow] (lh)--(rl);%
    
 	\draw[line width=2pt,dashed] (3.9,-3.0)--(3.9,5.0);%
	\draw[line width=2pt,dashed] (-2.4,-3.0)--(-2.4,5.0);%

\end{tikzpicture}
\begin{tikzpicture}[scale=1,
    shell2/.style={rectangle,minimum size=6mm,fill=red!50!white,draw=black},
    mixarrow/.style={<->,red,line width=2pt}]
    
    \foreach \x in {-2,...,-1}
        \draw[black!50!white] (-0.9,\x) -- (-0.1,\x)
          (1.6,\x) -- (2.4,\x);%
    \foreach \x in {1,3}
        \draw[black!50!white] (-0.9,\x) -- (-0.1,\x)
          (1.6,\x) -- (2.4,\x);%
    \draw[line width = 2pt] (-0.9,2) -- (-0.1,2)
          (1.6,2) -- (2.4,2)
          (-0.9,0) -- (-0.1,0)
          (1.6,0) -- (2.4,0);%
          
    \foreach \x in {-2,-1}
        \draw[->,very thick,black!50!white] (-0.5,\x-0.2) -- (-0.5,\x+0.2);%
    \foreach \x in {-2,-1}
        \draw[->,very thick,black!50!white] (2.0,\x-0.2) -- (2.0,\x+0.2);%
    \draw[->,line width=1.5pt] (-0.5,-0.2) -- (-0.5,0.2);%
    \draw[->,line width=1.5pt] (2.0,-0.2) -- (2.0,0.2);%
    
    \draw[loosely dotted, line width=1.2pt] 
        (-0.5,-3.0) -- (-0.5,-2.5)
        (2.0,-3.0) -- (2.0,-2.5)
        (-0.5,3.5) -- (-0.5,4.0)
        (2.0,3.5) -- (2.0,4.0);%
                      
    \node (lh) at (-1.5,0) {\large $\mL_{\kF}$};%
    \node (rh) at (3.0,0) {\large $\mR_{\kF}$};%
    \node [black!50!white] (ll) at (-1.5,1) {\large $\mL_{\kF+1}$};%
    \node [black!50!white] (rl) at (3.0,1) {\large $\mR_{\kF+1}$};%
    \node (ll1) at (-1.5,2) {\large $\mL_{\kF+2}$};%
    \node (rl1) at (3.0,2) {\large $\mR_{\kF+2}$};%

    \foreach \x in {1,...,2}
        \node [black!50!white] at (-1.5,-\x) {\large $\mL_{\kF-\x}$};%
    \foreach \x in {1,...,2}
        \node [black!50!white] at (3.0,-\x) {\large $\mR_{\kF-\x}$};%
    \foreach \x in {3,...,3}
        \node [black!50!white] at (-1.5,\x) {\large $\mL_{\kF+\x}$};%
    \foreach \x in {3,...,3}
        \node [black!50!white] at (3,\x) {\large $\mR_{\kF+\x}$};%

    \node [shell2]     (n+1)   at  (0.75,4.5)      {\Large ESWC with $p=n+1$};%

\begin{scope}[on background layer]
    \node [fill=red!20!white!,rounded corners=3mm,draw=black,fit=(lh) (rh) (ll1) (rl1)] {};%
\end{scope}%

    \node (ll1) at (-0.1,2) {};%
    \node (rl1) at (1.6,2) {};%
    \node (lh) at (-0.1,0) {};%
    \node (rh) at (1.6,0) {};%

    \draw[mixarrow] (ll1)--(rh);%
    \draw[mixarrow] (lh)--(rl1);%
    
\end{tikzpicture}
\end{adjustbox}
\caption{
\label{fig:mix}
Backward-mixing strategy used to generate various CDWs.
MO-pairs involved in mixing are bridged by red arrows. 
}
\end{figure}

\textit{Paradigm.---}
A 1DEG is constructed by confining $n$ electrons to a ring with radius $R$ and length $L = 2 \pi R$.
Thus the average electron density of such a system (also known as ``ringium'' in the literature \cite{QR12, Ringium13, gLDA14, NatRing16, SBLDA1D16}) is $\rho_0 = n/L=1/(2\rs)$, where $\rs$ is the so-called Wigner-Seitz radius.
Electrons interact Coulombically through the operator $r_{ij}^{-1}$, where $r_{ij} = R \sqrt{2 - 2 \cos(\theta_i - \theta_j)}$ is the across-the-ring distance between electrons $i$ and $j$, and $\theta_{i}$ is the angular displacement of an electron $i$.
For the sake of simplicity, we consider a curvilinear coordinate system henceforth, where $x_i = L \theta_i/(2\pi)$, and, without loss of generality, we set $L = 2\pi$ throughout this study.
We refer the readers to Ref.~\onlinecite{Ringium13} for more details about this paradigm.
Because the paramagnetic and ferromagnetic states are degenerate in strict 1D systems, we will consider only the spin-polarized electron gas from hereon \cite{QR12, Ringium13, 1DEG13, gLDA14, Wirium14, Lee11a, 1DChem15}.

\textit{Fermi fluid.---}
A FF is formed by occupying the $n$ lowest-energy plane waves (PWs) 
\begin{equation}
\label{eq:AO}
	\phi^{\text{FF}}_k(x) = \frac{\exp \left( i k x \right)}{\sqrt{L}},
\end{equation}
with $\abs{k} \leq \kF$, where $\kF = (n-1)/2 $ is the Fermi wave number and $\eF = \kF^2/2$ is the Fermi energy.
It has a (rotationally-invariant) uniform electron density
\begin{equation}
    \label{eq:rhoff}
    \pFF(x) = \sum_{\abs{k} \le \kF} \abs{\phi^\text{FF}_k(x)}^2 = \rho_0,
\end{equation}
and the density matrix is
\begin{equation}
\label{eq:qkFF}
	P_{k_1 k_2} = 
	\begin{cases} 
		\delta_{k_1 k_2},	&	\abs{k_1} \leq \kF, 
		\\
		0,				&	\abs{k_1} > k_\text{F},
	\end{cases}
\end{equation}
where $\delta_{k_1 k_2}$ is the Kronecker delta \cite{NISTbook}.
Thanks to its high symmetry, the HF energy is simply \cite{Ringium13}
\begin{equation}
\label{eq:eHF-1D}
	\eFF(\rs,n) = \tFF(\rs,n) + \vFF(\rs,n),
\end{equation}
where
\begin{subequations}
\begin{align}
	\tFF(\rs, n) = 	& \frac{1}{\rs^2} \qty( \frac{\pi^2}{24} \frac{n^2-1}{n^2} ),
				\\
	\vFF(\rs, n) = 	& \frac{1}{\rs} \qty( \frac{1}{2} - \frac{1}{8n^2} )  
    				 \qty[ \psi \qty(n+\frac{1}{2}) - \psi \qty(\frac{1}{2} ) ] - \frac{1}{4}
\end{align}
\end{subequations}
are the kinetic and potential energies respectively, and $\psi(x)$ is the digamma function \cite{NISTbook}.

\begin{figure}
\includegraphics[width=\linewidth]{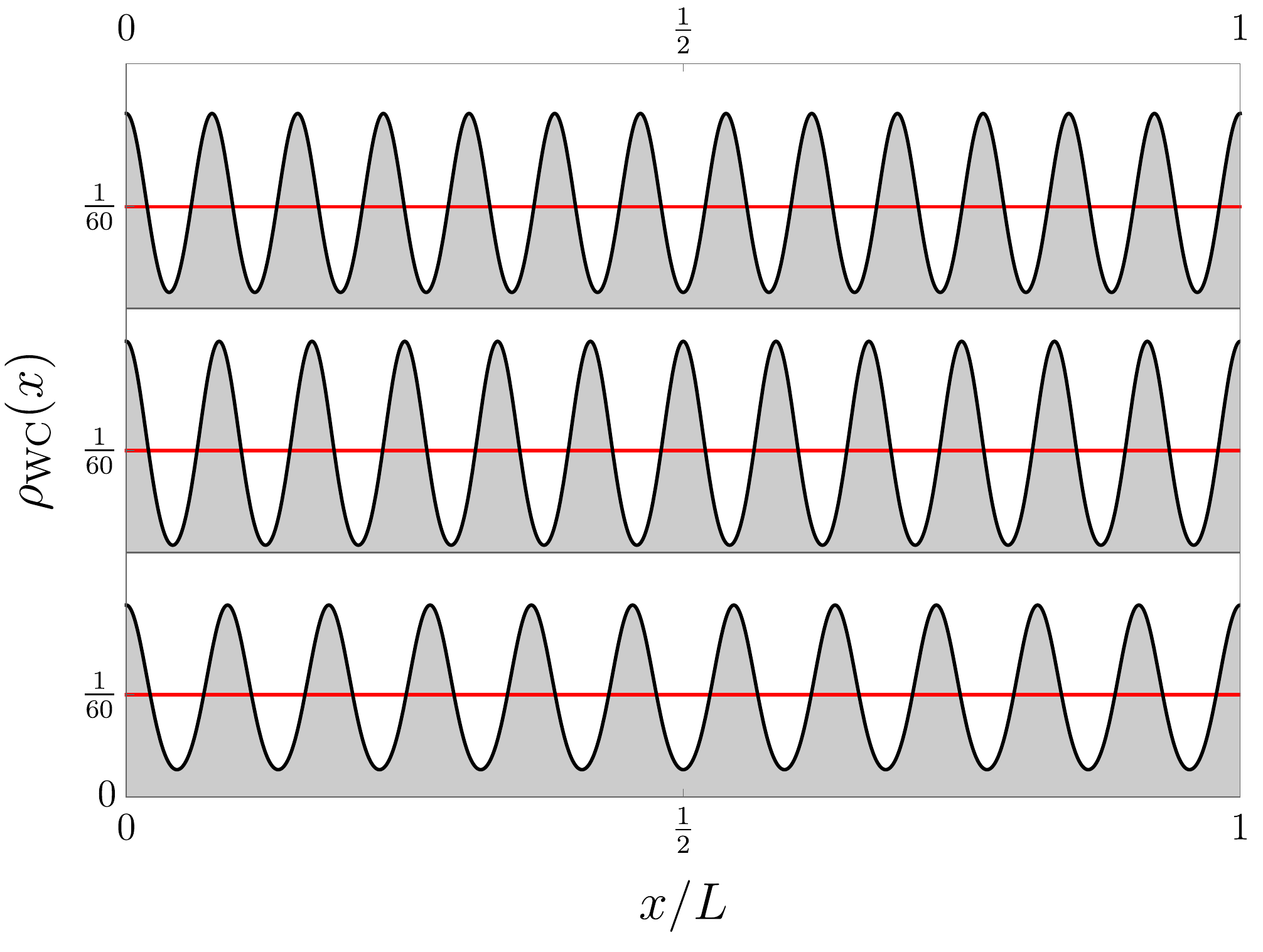}
\caption{
\label{fig:density}
WC densities for $n = 12$, $\rs = 30$ and $p = n - 1$ (bottom), $p = n$ (middle) and $p = n+1$ (top).
The FF density is represented as a solid red line for reference.}
\end{figure}

\textit{Symmetry-broken states.---}
In order to obtain symmetry-broken (SB) states, we have written a SCF HF program \cite{SzaboBook} using PWs of the form \eqref{eq:AO} with 
\begin{align}
	k = -\frac{M-1}{2}, \ldots, \frac{M-1}{2},
\end{align}
where $M$ is the total number of basis functions, which we have taken up to $M=400$ if required.
Basis functions with signed exponents are members of the ``left'' ($\mL$) or ``right'' ($\mR$) set, i.e.
\begin{equation}
\label{eq:leftright}
	\phi^{\text{FF}}_k(x) = 
		\begin{cases}
			\mL_k(x),	&	k < 0,
			\\
			\mR_k(x),	&	k > 0.
		\end{cases}
\end{equation}
This ``symmetry-broken'' HF (SBHF) program requires one- and two-electron integrals and they can be found in Ref.~\onlinecite{Ringium13}.

In general, the SCF procedure reliably returns the HF ground state solution, which is always either the FF or the GSWC \cite{SBLDA1D16}.
Consequently, in order to capture an ESWC, one requires a suitable guess density prior to starting the SCF process \cite{SzaboBook}.

\begin{figure*}
\includegraphics[height=0.37\textheight]{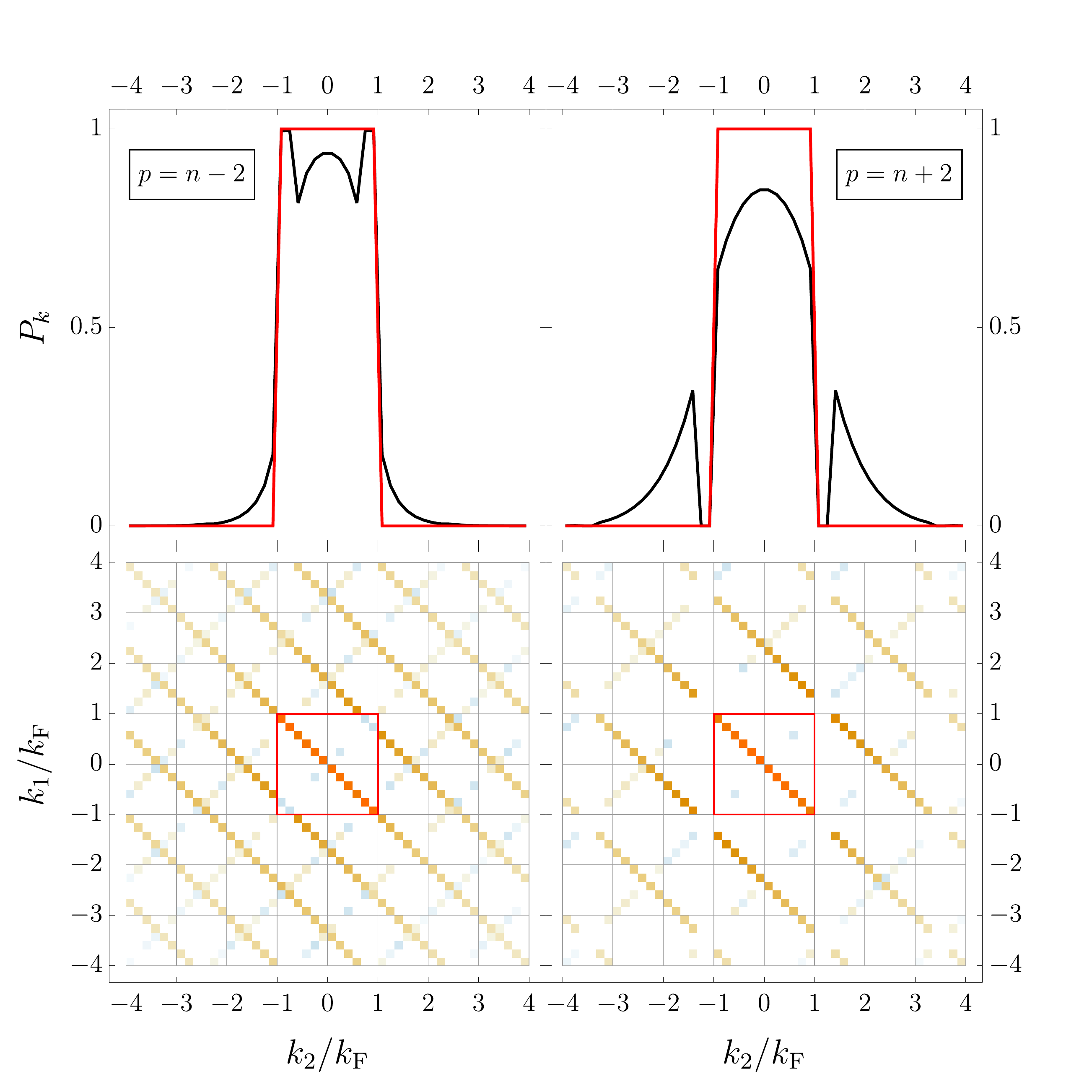}
\includegraphics[height=0.37\textheight]{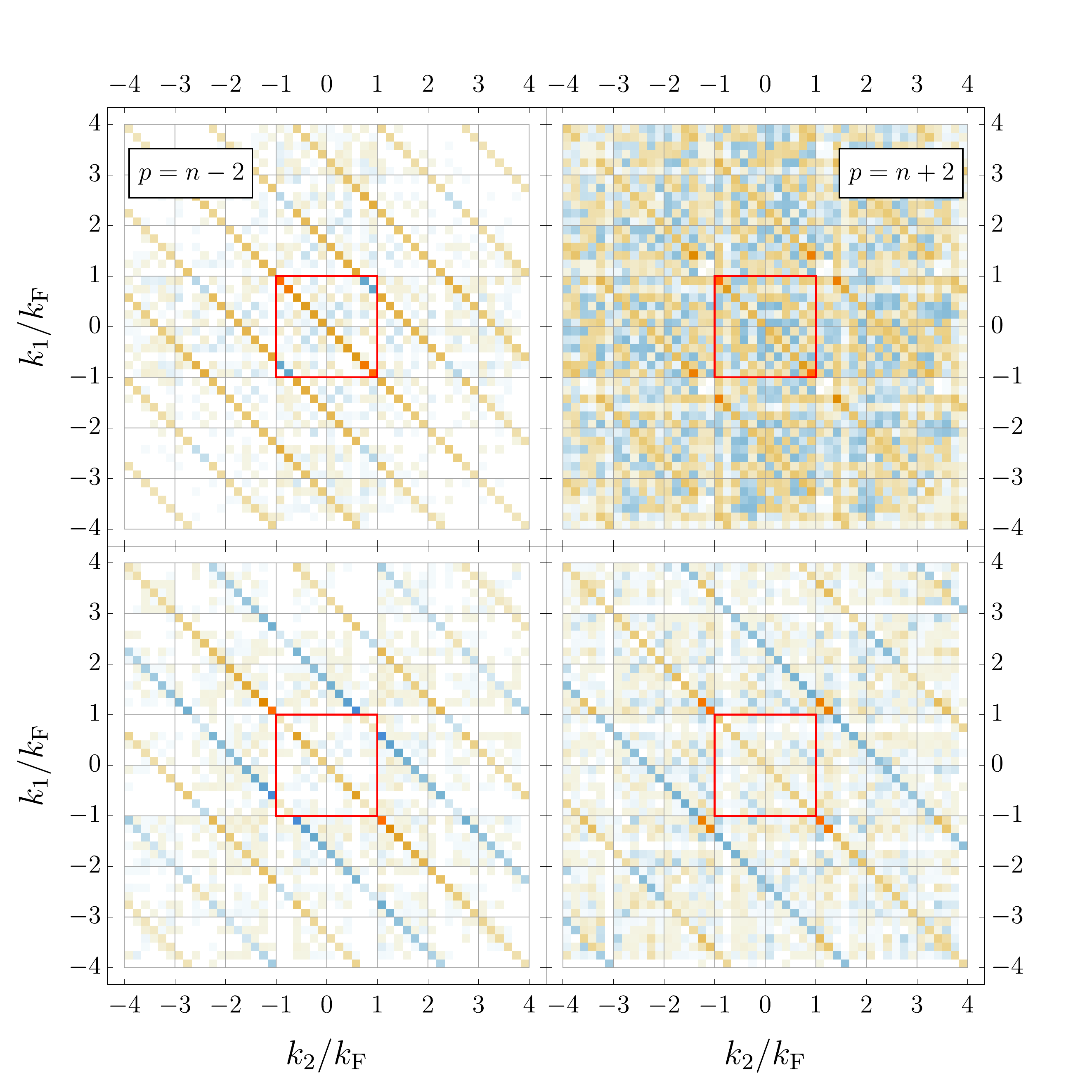}
\caption{
\label{fig:DM}
PW populations (top left), density matrix (bottom left), detachment (top right) and attachment (bottom right) matrices for $p=n-2$ (left) and $p=n+2$ (right) with $n=12$ and $\rs=30$.
The FF is represented as a solid red line for reference.}
\end{figure*}

\textit{Ground-state Wigner crystal.---}
One can, with little difficulty, generate a symmetry-valid guess density to obtain the commensurate $n$-peak GSWC.
As depicted in the central column of Fig.~\ref{fig:mix},  starting with the FF molecular orbitals (MOs), this is performed by cross-mixing the right and left highest-occupied MOs (HOMOs) with the opposed lowest-unoccupied MOs (LUMOs):
\begin{subequations}
\begin{align}
	\label{eq:wcmixR}
	\mR^{(0)}_{k_\text{F}}(x) & = \cos(\omega) \mR_{k_\text{F}}(x) + \sin(\omega) \mL_{k_\text{F} + 1}(x), 
	\\
	\label{eq:wcmixL}
	\mL^{(0)}_{k_\text{F}}(x) & = \cos(\omega) \mL_{k_\text{F}}(x) + \sin(\omega) \mR_{k_\text{F} + 1}(x),
\end{align}
\end{subequations}
where $\mL^{(0)}_{k_\text{F}}$ and $\mR^{(0)}_{k_\text{F}}$ are the post-mixing HOMOs  and $0 < \omega < \pi/2$ can be chosen to minimize the energy of the two-electron system composed by the MOs \eqref{eq:wcmixR} and 	\eqref{eq:wcmixL}.
Because this cross-mixing strategy is somewhat similar to the backward-scattering process in the Luttinger liquid model \cite{Haldane81}, we will refer to this as ``backward-mixing'' \footnote{Although we have taken one particular route to generate our CDWs, this choice is not unique. For example, one might also invoke ``forward-scattering'', as in the Luttinger liquid model \cite{VignaleBook}.}.

The initial density
\begin{equation}
\label{eq:rhowc}
    \pGSWC^{(0)}(x) 		= \frac{n}{L} + A \cos(n x) 
   					= \pFF(x) + \pCDW(x,n),
\end{equation}
(where $A = \pi^{-1} \sin 2\omega $) is simply a CDW, i.e.~a fluctuation of the FF uniform density \eqref{eq:rhoff} by a small sinusoidal modulation \cite{Frohlich54, Peierls55}.
Equation \ref{eq:rhowc} implies that one finds $n$ evenly-spaced peaks \textit{on} the Fermi surface, which corresponds exactly to the symmetry of the desired GSWC.

Generation of the CDW accompanies the appearance of band gaps at $\pm \kF$ \cite{Gruner88, Berlinsky79}.
Thus, one can liken the appearance of CDW to a ``nucleation'' event which, when carried through the SCF process, evolves into the WC, much like the growth of a crystal upon the addition of a ``seed'' to a supercooled liquid.
The symmetry of the seed (i.e.~the CDW) is paramount as it is sustained during crystal growth.
This observation affords special consideration, as Shore et al.~have suggested that CDWs may in fact account for the primordial phase in the growth of a WC \cite{Shore78} (see also Refs.~\cite{Sanders80, Perdew80}).
The present study confirms their conjecture.
Supplementary Material contains movies showing the growth of the initial CDW into a WC during the SCF process \footnote{See Supplemental Material at [URL will be inserted by publisher] for movies that illustrate the evolution of a CDW to a WC during the SCF process for $n = 12$ at $\rs = 30$.}.

A local stability analysis \cite{Seeger77} shows that, for $\rs > \rsSB(n)$, the GSWC is a genuine minimum of the HF equations while, for $\rs < \rsSB(n)$ the FF is the true HF ground state.
$\rsSB(n)$ slowly tends to zero for large $n$ as $(\ln n)^{-1}$ \cite{SBLDA1D16}, confirming Overhauser's prediction that, in the thermodynamic limit (i.e.~$n \to \infty$), it is always favorable to break the spatial symmetry \cite{Overhauser59, Overhauser62}.

\textit{Excited-state Wigner crystals.---}
Producing a $p$-peak CDW with the symmetry of an ESWC is simply a matter of selecting the right orbitals to mix.
For example, for $p>n$, one may mix the HOMOs with a pair of virtual MOs of higher angular momentum than the LUMOs as illustrated in the right column of Fig.~\ref{fig:mix}, where $p = n+1$. 
Similarly, one may reduce $p$ by backward-mixing the LUMOs with a pair of occupied MOs of lower angular momentum than the HOMOs (see left column of Fig.~\ref{fig:mix} where $p = n-1$).
In either case, the mixed orbitals must be separated by $\Delta k = 2 \kF + 1 + \delta $ (where $\delta = p - n$) to create a $p$-peak CDW:
\begin{equation}
\label{eq:rhoewc} 
    \pWC^{\text{(0)}}(x)	= \frac{n}{L} + A \cos(p x) 
					= \pFF(x) + \pCDW(x,p).
\end{equation}
Similarly to the GSWC, starting the SCF with the guess density \eqref{eq:rhoewc} will nucleate an ESWC of the desired symmetry, i.e.~possessing $p$ peaks separated by a distance $L/p$ (see Supplementary Material).
The electron density of the $(n-1)$- and $(n+1)$-peak ESWCs are represented in Fig.~\ref{fig:density}, where the $n$-peak GSWC is also reported.

The density matrix $P_{k_1 k_2} = \sum_i^\text{occ} c_{k_1 i} c_{k_2 i}$ (where $c_{k i}$ is the $k$th PW coefficient of the $i$th MO) and the PW population $P_{k} = \sum_{k_2}^M P_{k k_2}$ of ESWCs for $p = n \pm 2$ are represented in the left panel of Fig.~\ref{fig:DM}.
It is interesting to note that, for $p=n+2$, PWs with $\abs{k}=\kF+1$ and $\kF+2$ are unpopulated (i.e.~$P_k = 0$), as are their harmonics at $\abs{k}=(2q+1)(\kF+1)$ and $(2q+1)(\kF+1)+1$ (where $q \in \mathbb{N}^*$). 
For unsaturated ESWCs, sets of PWs from $\abs{k}=(2q+1)(\kF+1)$ to $\abs{k}=(2q+1)(\kF+1) + \delta -1$ are unpopulated.
For $p \to \infty$, one reaches the FF limit.

PW populations of supersaturated ESWCs are distinctly different to those just described.
Rather than depopulating certain PWs, others are instead highly populated (i.e.~$P_k \approx 1$).
For example, the ($n-2$)-peak ESWC (reported in Fig.~\ref{fig:DM}) has PWs with $\abs{k}=\kF$ and $\kF -1$ highly populated. 
Generally speaking, an ESWC with $p < n$ peaks has its $n-p$ highest PWs with $\abs{k} \le \kF$ highly occupied. 
Again, when $p \to 0$, one reaches the FF limit.

For $\rs > \rsSB(n,p)$, the $p$-peak ESWC is lower in energy than the FF.
Similarly to the GSWC, we have found that $\rsSB(n,p)$ goes to zero at the same $(\ln n)^{-1}$ asymptotic rate, which extends Overhauser's prediction to ESWCs.
Unsurprisingly, a local stability analysis shows that, even for $\rs > \rsSB(n,p)$, the ESWCs correspond to saddle points of the HF equations and are never true minima.
Moreover, the number of negative eigenvalues (i.e.~the order of the saddle point) increases with $\abs{\delta}$, and reaches its maximum for the FF.

\begin{figure}
\includegraphics[height=0.29\textheight]{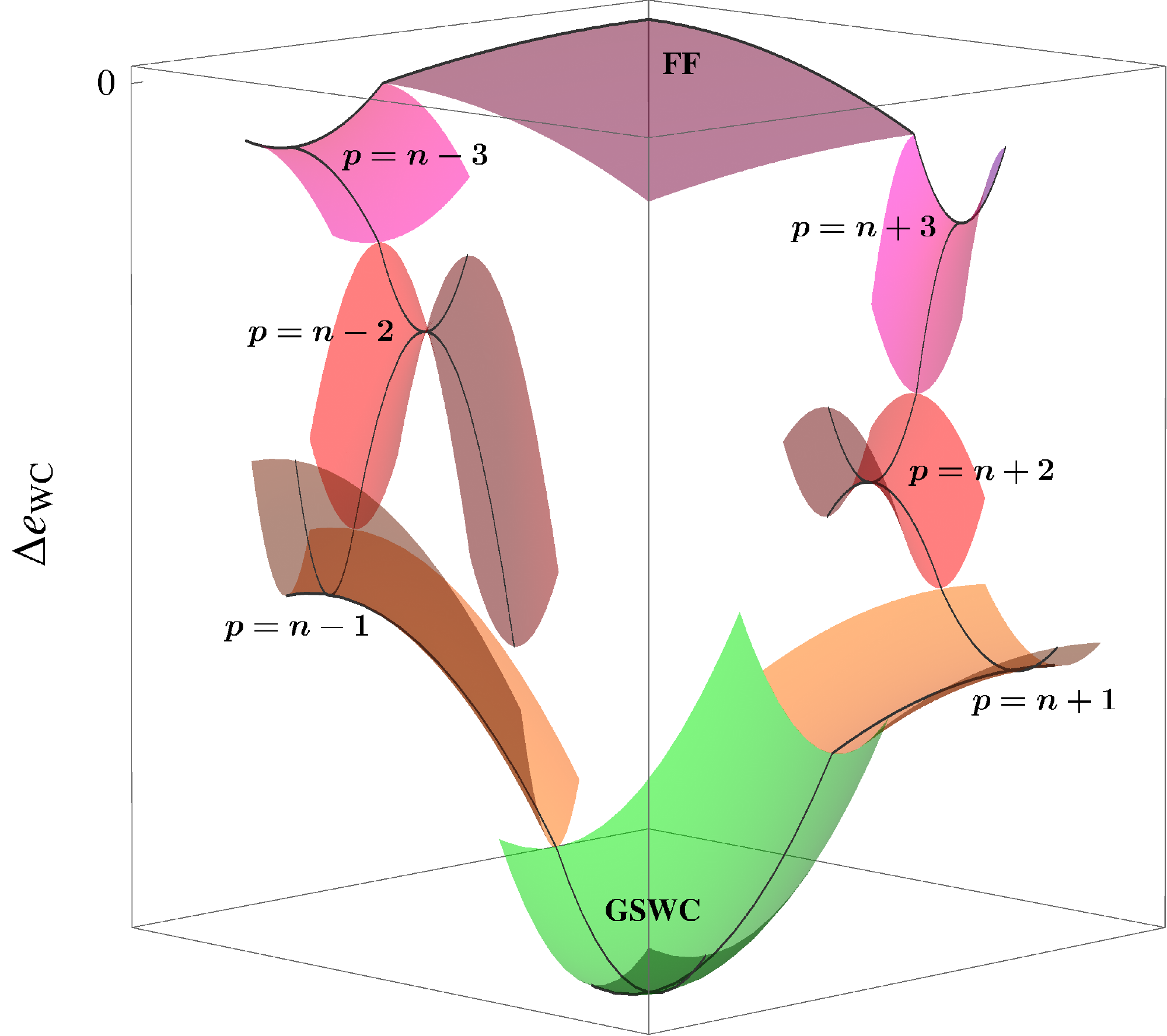}
\caption{
\label{fig:fountain}
Schematic representation of the HF manifold and its stationary points.
The GSWC is the true HF ground state, while ESWCs correspond to saddle points of increasing order.
}
\end{figure}

To further understand this, we have reported in Fig.~\ref{fig:DM} the detachment and attachment matrices \cite{Dreuw05} corresponding to the lowest negative eigenvalue for ESWCs with $p = n \pm 2$ (right panel).
For $p=n+2$, it shows that, to lower the energy, one needs to transfer electron density from PWs with $\abs{k} \lesssim \kF$ to the unpopulated PWs $\mL_{\kF+1}$ and $\mR_{\kF+1}$.
Similarly, for $p=n-2$, one needs to transfer electron density from the highly populated PWs $\mL_{\kF}$ and $\mR_{\kF}$ to PWs with $\abs{k} \gtrsim \kF +1$.
Other eigenvectors associated with negative eigenvalues yield a similar picture.
Starting from the FF, following the eigenvectors corresponding to these instabilities evidences that ESWCs of each variety are inter-connected with each other within the HF manifold, and ultimately reaches the GSWC.
This is schematically represented in Fig.~\ref{fig:fountain}.

\textit{Energetics.---}
As in previous work \cite{SBLDA1D16}, we measure stabilization energies defined as follows,
\begin{subequations}
\label{eq:sbhf}
\begin{align}
	\Delta \eWC(\rs,n,p) & = e_{\text{WC}}(\rs,n,p) - e_{\text{FF}}(\rs,n), 
	\\
	\Delta \tWC(\rs,n,p) & = t_{\text{WC}}(\rs,n,p) - t_{\text{FF}}(\rs,n), 
	\\
	\Delta \vWC(\rs,n,p) & = v_{\text{WC}}(\rs,n,p) - v_{\text{FF}}(\rs,n).
\end{align}
\end{subequations}
To extrapolate our results to the thermodynamic limit, we have employed the same procedure as in Ref.~\cite{SBLDA1D16}.

\begin{figure}
\includegraphics[width=\linewidth]{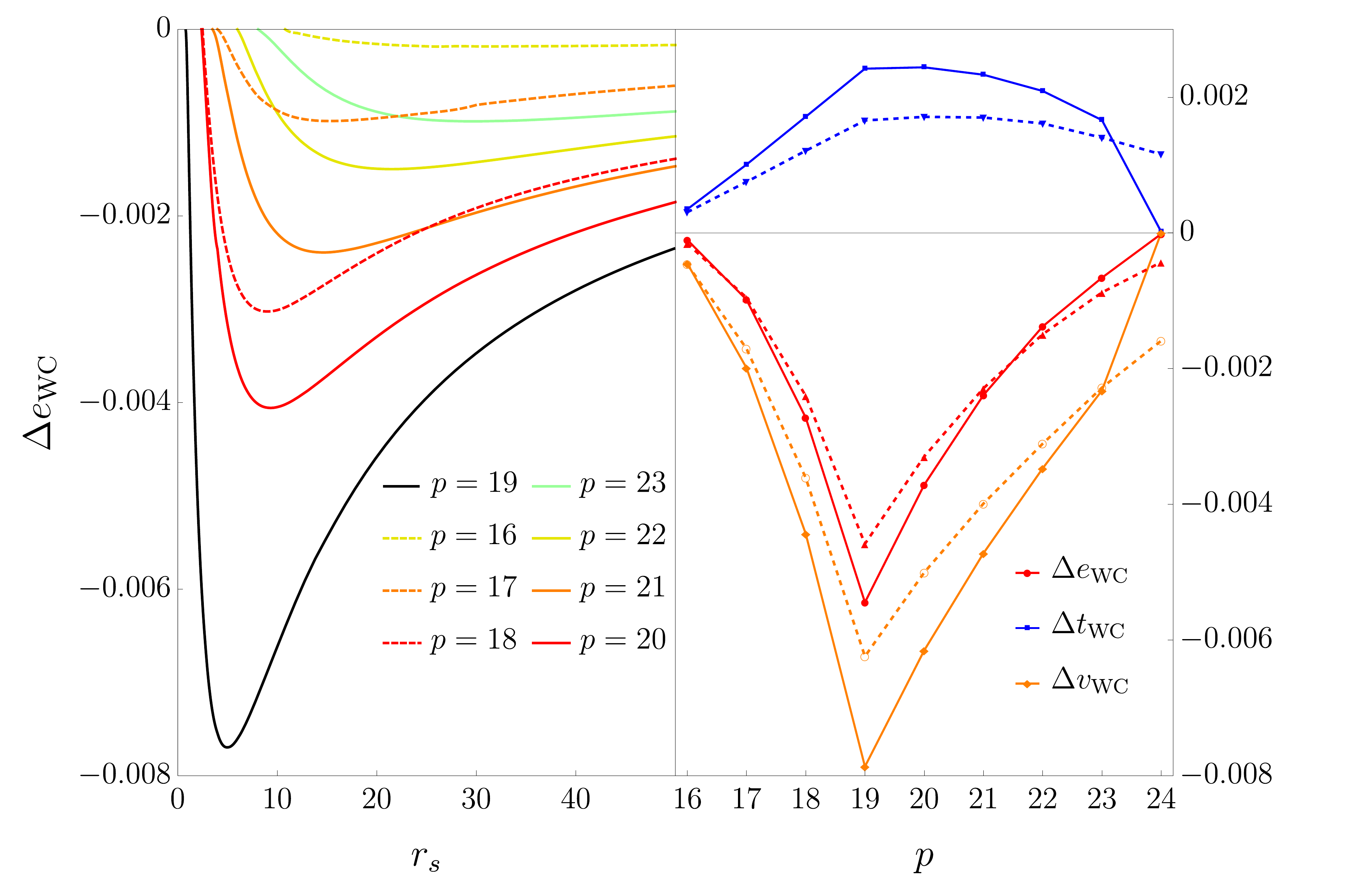}
\caption{
\label{fig:stabilisation}
Left: $\Delta e_{\text{WC}}$ as a function of $\rs$ for $n = 19$ and various $p$.
Right: $\Delta t_{\text{WC}}$ (blue), $\Delta v_{\text{WC}}$ (orange) and $\Delta e_{\text{WC}}$ (red) as a function of $p$ for $n=19$, $r_s=15$ (solid) and $r_s=20$ (dashed).}
\end{figure}

It is instructive to understand the energetics of ESWCs as functions of $\rs$ and $p$, especially given our claim that these are excited states.
To this end, we have computed $\Delta e_{\text{WC}}$ as a function of $\rs$ for $n=19$ and $16 \leq p \leq 23$.
The results are depicted in the left graph of Fig.~\ref{fig:stabilisation}.
For all densities, it is clear that successive removal or addition of peaks to the GSWC is adjoined to an increase in $\Delta e_{\text{WC}}$.
Furthermore, one finds that $\rsSB$ decreases by the same action.
Interestingly, the change incurred from decreasing $p$ is more significant than that resulting from its increase.

\begin{figure*}
\includegraphics[width=\linewidth]{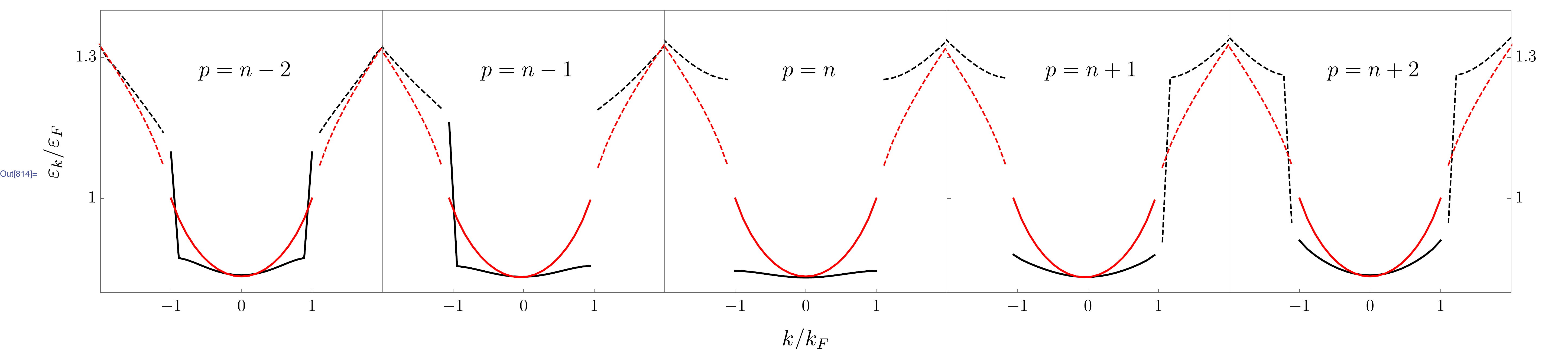}
\caption{
\label{fig:gaps}
MO energies $\eps_k$ as a function of $k$ for various WCs with $n=19$ and $r_s=15$. 
Occupied and virtual MOs are represented by solid and dashed lines, respectively. 
The FF MO energies are represented in red for reference.}
\end{figure*}

Figure \ref{fig:stabilisation} also evinces that the ESWC energy hierarchy does change with $\rs$.
For instance, the $18$-peak ESWC is lower in energy than the $21$-peak ESWC up to $r_s \approx 25$, after which the unsaturated ESWC is energetically favoured.
Furthermore, we find that these transitions only occur between a supersaturated crystal and an unsaturated crystal, and not between ESWCs of the same variety.
This indicates that of the two crystal species, the stabilization energy of a supersaturated ESWC decays most rapidly for large $r_s$, which correlates with the large-$\rs$ behavior of the WC energy \cite{Ringium13}:
\begin{equation}
	\eWC(\rs,n,p) = \frac{n^2}{p^2} \frac{\pi}{2n^2 \rs} \sum_{k=1}^{p-1} \frac{p-k}{\sin(k \pi/p)} + \order{\rs^{-3/2}}.
\end{equation}

Stabilization energies extrapolated in the thermodynamic limit for the $p=n \pm 1$ ESWCs and the $n$-peak GSWC are given in Table \ref{tab:limits}.
We find that $\Delta e_{\text{ESWC}}$ is always higher than $\Delta e_{\text{GSWC}}$.
Thus, these ESWCs are true to their name, as they never are the HF ground state.

As illustrated in the right graph of Fig.~\ref{fig:stabilisation}, a WC will only form over a FF if the gain in potential energy ($\Delta \vWC$) is larger than the loss of kinetic energy ($\Delta \tWC$). 
For $n=19$ and $r_s=15$, the GSWC sustains the largest decrease in $\vWC$, at the cost of a much smaller increase in $\tWC$. 
Thus, as expected, $\Delta e_{\text{WC}}$ reaches its maximum when $p=n$.
It also neatly reveals the asymmetry in the character of the supersaturated and unsaturated crystals.
Clearly, $\Delta t_{\text{WC}}$ and $\Delta v_{\text{WC}}$ diminish most rapidly with the removal of peaks.
Moreover, both $\Delta t_{\text{WC}}$ and $\Delta v_{\text{WC}}$ approach zero with the addition or removal of peaks from the GSWC, while solutions where $p > 24$ or $p < 16$, are unstable to the FF at this density.
We have found that the stabilization energy is marginal when $\abs{\delta}$ is large, corresponding to a very small fluctuation of the FF density. 
However, it is easier to create ESWCs with large $\abs{\delta}$ when $n$ increases.
\begin{table}
\caption{
\label{tab:limits}
$\Delta e_{\text{WC}}$ (in millihartrees) for various WCs in the thermodynamic limit. 
}
\begin{ruledtabular}
	\begin{tabular}{lcccc}
		$p$		&   \mc{4}{c}{$\rs$}  \\
\cline{2-5}
				&	2  		&	5   		&	10  		&	15		\\
		\hline
		$n-1$	&	$-4.89$	&   $-7.24$	&	$-6.22$	&	$-5.00$	\\
		$n$		&	$-5.94$	&   $-8.00$	&	$-6.76$	&	$-5.54$	\\
		$n+1$	&	$-4.93$	&   $-7.37$	&	$-6.43$	&	$-5.21$	\\
	\end{tabular}
\end{ruledtabular}
\end{table}

\textit{MO energies.---}
To explore the electronic properties of ESWCs, we have studied the MO energies $\eps_k$ and, in particular, their HOMO-LUMO gap. 
The results are reported in Fig.~\ref{fig:gaps} for $n = 19$ and $r_s = 15$.
In the FF, the HOMO-LUMO gap is small \footnote{In the thermodynamic limit, the HOMO-LUMO gap of the FF is exactly zero.}, giving rise to a metallic character, while the GSWC exhibits a large gap at $k=\pm \kF$.
Depending on the magnitude of the gap, the GSWC behaves either as a semi-conductor or an insulator.
However, for the sake of discussion, we will assume that large gaps are always insulating.

Interestingly, for $p=n+1$, one virtual MO (with positive or negative momentum) is lowered to the region of the occupied set.
This induces an overall destabilization of the occupied MOs compared to the GSWC.
The gap becomes asymmetric with one half resembling the GSWC, and the other resembling the FF.
For $p=n-1$, the opposite scenario is drawn.
Here, one occupied MO is raised in energy to the neighborhood of the unoccupied set, and so the gap on this side disappears.
Addition or removal of two or more peaks results in the loss of gaps altogether, as in both examples, the energies of the HOMOs and LUMOs become similar (see leftmost and rightmost graphs in Fig.~\ref{fig:gaps}).

In the thermodynamic limit, ESWCs would possess unusual conductivity properties given the form of their band gap.
For instance, in the $(n-1)$-peak ESWC, the high-energy electron is only conductible in the direction of their initial momentum.
In this sense, the $p=n-1$ ESWC is an anisotropic or ``chiral'' conductor (in reference to the chiral Luttinger liquid model \cite{Chang03}).
Similarly, we can infer that the $(n-2)$-peak ESWC withholds the properties of an isotropic or achiral conductor, since it possesses conductive electrons of both positive and negative momenta.
In contrast, the unsaturated ESWCs would exhibit this tendency as $\delta$ grows large, while behaving as insulators when $\delta$ is small.

\textit{Acknowledgments.---}
P.F.L.~thanks the Australian Research Council for a Discovery Early Career Researcher Award (DE130101441) and a Discovery Project grant (DP140104071), and the NCI National Facility for generous grants of supercomputer time.

\end{document}